\documentclass{article}
\usepackage{layouts}
\usepackage{makecell}

\usepackage{lineno}
\nolinenumbers
\usepackage{pdflscape}
\usepackage{array}
\usepackage[longtable]{multirow}
\usepackage{longtable, pbox, rotating, hvfloat}
\usepackage{threeparttable}
\usepackage{layouts}
\usepackage{booktabs}
\usepackage{breakcites}
\usepackage{multicol}
\usepackage{gensymb}
\usepackage{notoccite}
\usepackage{soul}
\usepackage[round]{natbib}
\usepackage{amsmath, amssymb}

\usepackage[table, dvipsnames]{xcolor}

\usepackage{tabularray}
\usepackage{epigraph}
\usepackage{fancyhdr}
\usepackage[colorlinks=true,citecolor=red,linkcolor=black]{hyperref}
\usepackage[Lenny]{fncychap}
\usepackage{scalerel}
\usepackage{pgfplots}
\usepackage{wrapfig}
\usepackage[font=small,labelfont=bf]{caption}
\usepackage[notransparent]{svg}
\usepackage[margin=.9in]{geometry}
\usepackage{breakcites}
\usepackage{amsmath, amssymb}
\definecolor{lightgray}{gray}{0.983}

\usepackage{epigraph}
\usepackage{fancyhdr}
\usepackage[colorlinks=true,citecolor=red,linkcolor=black]{hyperref}
\usepackage[Lenny]{fncychap}
\usepackage{scalerel}
\usepackage{pgfplots}
\usepackage{wrapfig}

\usepackage{enumitem}
\usepackage{verbatim}
\usepackage{float}
\usepackage{graphicx , array}
\usepackage{adjustbox}

\usepackage{tikz}
\usepackage[graphicx]{realboxes}
\usepackage{pgfornament}

\usepgfplotslibrary{groupplots,dateplot}
\pgfplotsset{compat=newest}
\usetikzlibrary{positioning, patterns,calc, arrows}

\usetikzlibrary{positioning,arrows,calc}

\usepackage{subcaption}
\usepackage{makecell}

\usepackage{enumitem}
\usepackage{verbatim}
\usepackage{float}
\usepackage{graphicx}
\usepackage{adjustbox}

\usepackage{tikz}
\usepackage[graphicx]{realboxes}
\usepackage{pgfornament}

\usepgfplotslibrary{groupplots,dateplot}
\pgfplotsset{compat=newest}

\usetikzlibrary{positioning, patterns,calc, arrows}

\definecolor{navyblue}{rgb}{0.0, 0.0, 0.5}

\usetikzlibrary{positioning,arrows,calc}

\usepackage{xspace}

\newcommand{\carbonic}{H\textsubscript{2}CO\textsubscript{3}\xspace}
\newcommand{\bicarbonate}{HCO\textsubscript{3}\ensuremath{^-}\xspace}
\newcommand{\carbonate}{CO\textsubscript{3}\ensuremath{^{2-}}\xspace}

\newcommand{\proton}{H\textsuperscript{+}\xspace}
\newcommand{\COtwo}{CO\textsubscript{2}\xspace}
\newcommand{\COtwoaq}{CO\textsubscript{2(aq)}\xspace}
\newcommand{\Otwo}{O\textsubscript{2}\xspace}
\newcommand{\Ntwo}{N\textsubscript{2}\xspace}
\newcommand{\SOtwo}{SO\textsubscript{2}\xspace}
\newcommand{\HtwoO}{H\textsubscript{2}O\xspace}
\newcommand{\HtwoS}{H\textsubscript{2}S\xspace}
\newcommand{\MgCltwo}{MgCl\textsubscript{2}\xspace}
\newcommand{\CaCltwo}{CaCl\textsubscript{2}\xspace}
\newcommand{\NaSOfour}{Na\textsubscript{2}SO\textsubscript{4}\xspace}
\newcommand{\sodium}{Na\textsuperscript{+}\xspace}

\newcommand{\calcium}{Ca\textsuperscript{2+}\xspace}
\newcommand{\magnesium}{Mg\textsuperscript{2+}\xspace}

\newcommand{\KCl}{KCl\xspace}

\usepackage{bibentry}

\begin{document}

\begin{center}

	\rule{\linewidth}{3pt}
    \Large
Enhanced Prediction of \COtwo Solubility under Geological Conditions for CCUS via Improved Pitzer Parameters and Physics-Informed Machine Learning
	\rule{\linewidth}{1pt}
\end{center}

Abdeldjalil Latrach\textsuperscript{1*}, Lily Jackson\textsuperscript{2}, Minou Rabiei\textsuperscript{1}

\textsuperscript{1} Energy and Petroleum Engineering Department, University of Wyoming\\
\textsuperscript{2} University of Texas at El Paso\\
\textsuperscript{*} Corresponding author: \texttt{alatrach@uwyo.edu}

\leftskip=1cm
\rightskip=1cm

\paragraph{Abstract} The solubility of \COtwo in formation brines plays a critical role in the efficiency of carbon capture and storage (CCS) operations. It is strongly influenced by pressure, temperature, and brine composition. Various experimental studies and modeling approaches have been developed to estimate CO2 solubility under wide ranges of pressure, temperature, and salinities. This work makes three key contributions. First, we present an extensive literature review of experimental, theoretical, and simulation-based approaches for measuring and predicting \COtwo solubility across a wide range of conditions and also a discussion of how the different parameters affect solubility. Second, we introduce an improved set of temperature-dependent Pitzer interaction parameters, yielding up to a 76\% reduction in average absolute deviation compared to conventional values in the geochemical simulation software PHREEQC. Third, we develop a physics-informed machine learning model that integrates thermodynamic intuition with data-driven learning, achieving a 14\% reduction in prediction error over the state-of-the-art and up to 40\% improvement at high salinities. Together, these advances provide a robust and accurate framework for predicting \COtwo solubility, supporting more reliable CCS design and deployment.


\leftskip=0pt\rightskip=0pt

\begin{multicols}{2}\raggedcolumns

\section{Introduction}
\begin{figure}[H]
\centering
\includegraphics[width=0.9\linewidth]{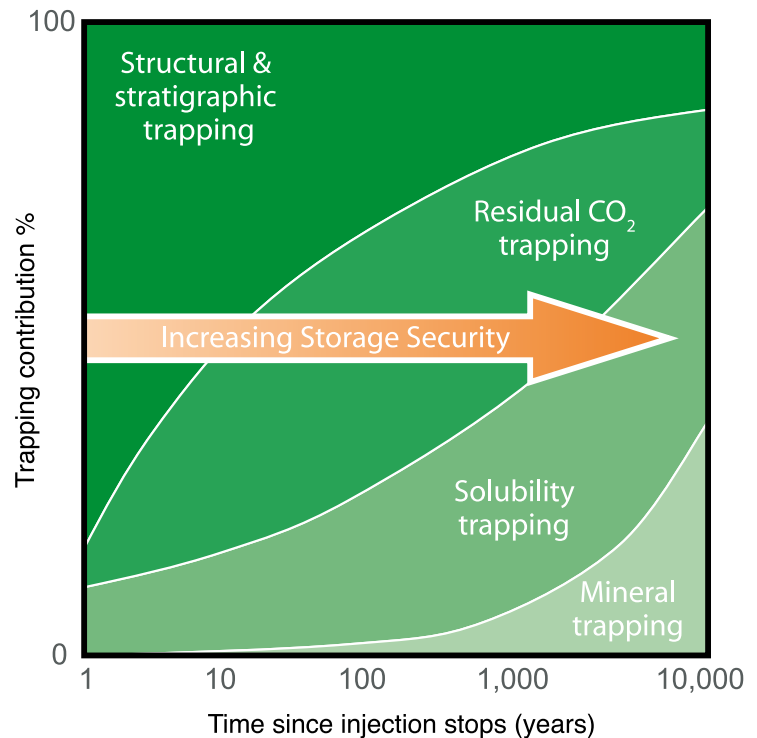}
\caption{Time-scale of different \COtwo trapping mechanisms and the change in their contribution over time \citep{metz2005ipcc}.}
\label{figure:trapping_mechanisms}
\end{figure}

Greenhouse gas (GHG) emissions are the primary driver of climate change and global warming. Carbon capture, utilization, and storage (CCUS) offers a promising strategy to mitigate these emissions by reducing the release of carbon dioxide (\COtwo) from industrial, transportation, and other sources into the atmosphere. This is achieved by permanently storing \COtwo in geological formations \citep{Davoodi2023}. Suitable geological formations for \COtwo sequestration include saline aquifers, depleted oil and gas reservoirs, and basalt formations \citep{raza2022}. The sequestration process involves different trapping mechanisms, which vary depending on the chosen formation. This study specifically focuses on \COtwo sequestration in saline aquifers. When \COtwo is injected into a saline aquifer, four main trapping mechanisms are active: 1) hydrodynamic, 2) solubility, 3) residual, and 4) mineral trapping \citep{DeSilva2015}. Each mechanism contributes to the total storage capacity at different rates and over varying timescales, as illustrated in Figure \ref{figure:trapping_mechanisms}.

This study primarily focuses on solubility trapping, which refers to the dissolution of \COtwo into the formation brine until thermodynamic equilibrium. This mechanism is crucial for two main reasons: it directly increases the total \COtwo storage capacity, and it acts as a precursor to mineral trapping, which involves geochemical reactions in the aqueous phase. Consequently, accurately estimating the amount of \COtwo that can dissolve in formation brine is vital for successful CCUS projects.

\COtwo solubility in brines is primarily affected by three factors: 1) temperature, 2) pressure, and 3) the concentration and types of dissolved aqueous species. Impurities in the \COtwo stream, originating from the capture source and technology, also influence solubility, a topic discussed later in this paper. These influencing factors vary significantly between formations due to differences in depth, temperature gradients, and reservoir mineralogy \citep{Tang2015}. Therefore, precise measurement or estimation of \COtwo solubility under reservoir conditions is essential for the design and implementation of CCUS projects.

\COtwo solubility can be measured or estimated using laboratory experiments, physical models, or numerical simulations. While experimental studies are generally the most accurate, they have increasingly focused on simple synthetic brines (e.g., NaCl, \CaCltwo, \MgCltwo, KCl, and \NaSOfour) over the past few decades. Although these simplified conditions may not fully replicate complex real formation brines containing numerous aqueous species and trace elements, they offer valuable data for understanding the impact of individual species on solubility in a controlled setting. Despite their inherent limitations in accuracy, physical models and numerical simulations are critical because they can be incorporated into more complex compositional reservoir simulations. Conversely, experimental data is indispensable for calibrating and validating both physical and simulation models.

To address the limitations of current models and enhance the accuracy of \COtwo solubility prediction, this work provides a comprehensive review of experimental studies, including various setups and approaches. We then discuss different physical models and their development, followed by an overview of numerical simulations. Next, we analyze \COtwo solubility in brines and its controlling factors: pressure, temperature, water chemistry, and the presence of impurities. A key contribution of this paper is the presentation of a new set of temperature-dependent Pitzer interaction parameters, fitted using experimental data from the literature. Finally, we introduce a physics-informed machine learning model developed to estimate \COtwo solubility in brines across a wide range of conditions.
\section{A review of experimental studies and modeling approaches}
This section aims to be a comprehensive discussion of the different methods for measuring and estimating \COtwo solubility in brines. First, we examine experimental studies of \COtwo solubility in pure water, synthetic, and formation brines, along with their common setups. We then discusses key physical models and numerical simulation approaches.

\subsection{Experimental studies}
Experimental studies on \COtwo solubility in brines primarily use either analytical or synthetic methodologies, with modern advancements employing in-situ spectroscopic techniques. Analytical methods typically involve equilibrating \COtwo with the brine in a high-pressure cell at precisely controlled temperatures and pressures, followed by sampling the liquid phase \citep{Messabeb2017,Liu2011}. The dissolved \COtwo is then quantified through post-sampling analyses, such as conductometric titration \citep{Messabeb2017}, gravimetric determination of absorbed \COtwo after freezing out or absorption \citep{Bamberger2000}, or volumetric measurement of evolved gas \citep{Liu2011}. Synthetic methods, on the other hand, often utilize high-pressure view cells where known amounts of \COtwo and brine are introduced, and solubility is determined by observing the specific pressure or temperature conditions at which a phase boundary (e.g., first bubble) appears \citep{Kamps2007,Rumpf1993}. Contemporary research also increasingly employs in-situ Raman spectroscopy within high-pressure optical cells (e.g., fused silica capillary cells) to directly quantify dissolved \COtwo by correlating its characteristic Raman peak intensity or height ratios (e.g., \COtwo Fermi dyad relative to the \HtwoO O-H stretching band) with concentration, a relationship typically established through prior calibration with homogeneous solutions of known composition \citep{Guo2014,Wang2019}. Across these diverse setups, common critical elements include systems for precise temperature and pressure control, stirring mechanisms to ensure thermodynamic equilibrium, and accurate sensors for monitoring experimental conditions. Table \ref{table:studies} summarizes some of experimental studies of \COtwo solubility in pure water, synthetic, and formation brines that are used later in this study. There is a large number of experimental studies, and this is by no means an exhaustive list, however, we focused on including a wide variety of brines and temperature/pressure ranges.

\begin{table*}[]
\centering
\caption{Experimental studies of \COtwo solubility in pure water and brines}
\label{table:studies}
\begin{threeparttable}

\begin{tabular}{cccccc}
	\hline
\textbf{Study} & \textbf{Pressure (MPa)} & \textbf{Temperature (K)} & \textbf{System (molality range)} &  \\
	\hline
\cite{Rumpf1993}	& 	0--10		&	313.15--433.15	&	\NaSOfour (0.998--2.010)  		\\
\cite{Rumpf1994}	&	1.059--5.798	&	323.16--323.18 & Pure water	\\
				&	0.151--9.642	&	313.14--433.15 & NaCl (3.997--5.99)	\\
					&	0--10		&	313.13--433.16	&	(NH\textsubscript{4})\textsubscript{2}SO\textsubscript{4} 	\\
\cite{Bamberger2000}	& 4.05--14.11	&	323.20--353.10	& Pure water\\ 
\cite{Spycher2003}	&	0--60		&	285.15--373.15	&	Pure water\\
\cite{Chapoy2004}	&	0.19--9.33	&	274.14--351.15	&	Pure water\\
\cite{Bermejo2005}	&	1.55--8.34	&	296.73--369.65 & Pure water \\
					&	1.98--13.11	&	286.98--368.81 & \NaSOfour (0.25--1.0)\\
\cite{Kamps2007}	&	0.3361--9.395 &	313.1--433.1 &	KCl (1.994--4.05)\\
					&	0.2672--9.237&	313.2--393.2&	K\textsubscript{2}CO\textsubscript{3} (0.4272--1.7125)\\
\cite{Yan2011}		&	5.0--40.0	&	323.2--413.2	&	Pure water\\
					&	5.0--40.0	&	323.2--413.2	&	NaCl (1, 5)\\
\cite{Liu2011}		& 2.08--15.99	&	308.15--328.15	&	Pure water	\\
					&	2.06--15.84	&	308.15--328.15	&	NaCl:KCl:\CaCltwo (0.290:0.227:0.153)	\\
					&	2.37--15.68	&	308.15--328.15	&	NaCl:KCl:\CaCltwo (0.590:0.463:0.311)	\\
					&	1.34--15.56	&	308.15--328.15	&	NaCl:KCl:\CaCltwo (0.856:0.671:0.451)	\\
					&	2.10--15.83	&	318.15	&	NaCl (1.901)	\\
					&	2.09--15.81	&	318.15	&	KCl (1.490)	\\
					&	2.09--15.86	&	318.15	&	\CaCltwo (1.001)	\\
					&	2.48--15.99	&	318.15	&	NaCl:KCl (0.901:0.706)	\\
					&	2.48--15.99	&	318.15	&	NaCl:\CaCltwo (0.901:0.706)	\\
					&	2.48--15.99	&	318.15	&	KCl:\CaCltwo (0.901:0.706)	\\
\cite{Lucile2012}	&	0.54--5.14	&	298.15--393.15	&	Pure water	\\
\cite{Tong2013}		&	1.53--37.99	&	309.61--424.64	&	\CaCltwo (1, 3, 5) \\
					&	1.25--31.71	&	309.58--424.68	&	\MgCltwo (1, 3, 5)\\
\cite{Guo2014} 		&	10--120		&	273.15--533.15& Pure water\\
\cite{Zhao2015}		&	15			&	323.15--423.15	&	Pure water \\
					&	15			&	323.15--423.15	&	NaCl (1--6) \\
\cite{Zhao2015b} 	&	15			&	323--423		&	\CaCltwo (0.333--0.667)\\
				 	&	15			&	323--423		&	\NaSOfour (1.67--2.0)\\
				 	&	15			&	323--423		&	\MgCltwo (0.333--0.667)\\
					 &	15			&	323--423		&	KCl (0.50--1.00)\\
\cite{Zhao2015c}	& 5.05--20.07	&	328				&	Natural Mt. Simon brine\tnote{*}  \\
					&	10--17.5			&		323--423			&	Synthetic Mt. Simon brine 1\tnote{\dag}\\
					&		10--17.5		&		323--423			&	Synthetic Mt. Simon brine 2\tnote{\ddag}\\
					&		10--17.5		&		323--423			&	Synthetic Antrim Shale brine 1\tnote{\dag} \\
					&		10--17.5		&		323--423			&	Synthetic Antrim Shale brine 2\tnote{\ddag}\\
\cite{Messabeb2017}	&5.05--20& 323.15--423.15 & Pure water \\
					&	5.01--20.04&323.15--423.15 	& \CaCltwo (1--6)\\
\cite{Wang2019}		&	3.0--30.0	&	303.15--353.15	&	Pure water\\
&	3.0--30.0	&	303.15--353.15	&	NaCl (1, 2, 3)\\
\cite{dosSantos2020}	& 1.6--20.15	&	303.15--423.15 & Pure water \\
						& 1.6--20.17	&	303.15--423.15 & \NaSOfour (1, 1.998) \\\hline

\end{tabular}
\begin{tablenotes}
	\item[*] The study used a natural formation brine of complex chemistry, refer to the original study for its composition.
	\item[\dag] These are synthetic brines that are intended to be as close as possible to natural brines.
	\item[\ddag] These are NaCl and CaCl\textsubscript{2} brines that have the same ionic strength as the natural brines.
	\end{tablenotes}
	\end{threeparttable}
\end{table*}

Experimental data was compiled from studies in Table \ref{table:studies}. This data was then carefully and thoroughly checked and verified to ensure its quality. During this quality control step, we plotted the data and we were able to identify points that are obvious outliers, or those with excessive noise. The verified dataset was then used to fit new Pitzer parameters and train the machine learning model in subsequent sections.

\subsection{Physics-based modeling}
Physics-based models, or physical models, are models based on fundamental physical principles and thermodynamic laws. These models aim to describe the behavior of the system using established physical laws and employing frameworks like equations of state (EOS), activity models, or specialized theories (e.g., Debye-Hückel, Pitzer theory, scaled-particle theory) to account for non-ideal solution behavior. Their development involves formulating governing equations derived from these principles. These models often incorporate adjustable parameters that are then fitted to experimental data.

An early effort by \citet{Li1986} introduced a model that combined a cubic EOS (Peng-Robinson) for the gas and oil phases with Henry's Law for describing gas solubility in the aqueous phase. Scaled-Particle Theory (SPT) was used to physically account for the presence of salts, specifically NaCl, and subsequently modify the Henry's constant. SPT, which calculates the work needed to create a cavity within the solvent for a solute molecule and the interaction energy between solute and solvent, provided a physically intuitive---but simplistic---means to incorporate salt effects. \citet{Harvey1989} developed a more direct integration of ionic effects into a conventional non-electrolyte equation of state. Their approach modified Born's equation to describe the charging of ions and utilized the Mean Spherical Approximation (MSA) for ion-ion interactions and salt/solvent parameter was calibrated using osmotic-coefficient data at room temperature. In parallel, \citet{Zuo1991} explored the feasibility of extending the Patel--Teja (PT) Equation of State to manage phase equilibrium calculations for high-pressure electrolyte solutions. Their key contribution was the incorporation of a Debye-Hückel electrostatic contribution term into the PT EOS, thereby accounting for long-range interionic forces while preserving the cubic form of the equation. Binary interaction parameters were determined by fitting a wide array of experimental data, encompassing water-salt interactions (from osmotic coefficients), gas-water (from vapor-liquid equilibrium data), and salt-gas pairs (from low-pressure gas solubility data). A distinct approach was presented by \citet{Sorensen2002}, who extended Soave--Redlich--Kwong (SRK) EOS to model gas solubility in formation water containing diverse salts such as NaCl, KCl, and \CaCltwo. Their model treats these salts as hypothetical components endowed with hypothetical critical properties. The Huron--Vidal (HV) mixing rule was then applied to gas-water, water-salt, and gas-salt binaries. This methodology allowed the utilization of conventional EOS routines, simplifying the computational process. 

The seminal work of \citet{Duan2003} is arguably the most recognized model of this sort with validity within a long range of pressures, temperatures, and salinities. The work is rooted in thermodynamic principles and relies on the balance of chemical potentials of \COtwo in the vapor and liquid phases at equilibrium (Equation \ref{eq:chem_pot_equ}). 

\begin{equation}
\mu_{\textrm{\COtwo}}^{\textrm{l}} = \mu_{\textrm{\COtwo}}^{\textrm{v}}
\label{eq:chem_pot_equ}
\end{equation}

Chemical potentials in the vapor/gas (Equation \ref{eq:chem_pot_v}) and liquid/aqueous (Equation \ref{eq:chem_pot_l}) phases are then expressed in terms of standard chemical potentials in those phases ($\mu_{\textrm{\COtwo}}^{\textrm{v}(0)}$ and $\mu_{\textrm{\COtwo}}^{\textrm{l}(0)}$), fugacity ($f$) in the vapor phase, and activity ($a$) in the liquid phase, leading to the following equations:

\begin{align}
	\label{eq:chem_pot_l}\mu_{\textrm{\COtwo}}^{\textrm{l}} &= \mu_{\textrm{\COtwo}}^{\textrm{l(0)}} + RT \ln a_{\textrm{\COtwo}}(T, P, m) \\&= \mu_{\textrm{\COtwo}}^{\textrm{l(0)}}  +  RT \ln m_{\textrm{\COtwo}} \nonumber\\ &+RT\ln \gamma_{\textrm{\COtwo}}(T, P, m)\nonumber
\end{align}

\begin{align}
\label{eq:chem_pot_v}\mu_{\textrm{\COtwo}}^{\textrm{v}} &= \mu_{\textrm{\COtwo}}^{\textrm{v(0)}} + RT \ln f_{\textrm{\COtwo}}(T, P, y) \\\nonumber &= \mu_{\textrm{\COtwo}}^{\textrm{v(0)}}  +  RT \ln y_{\textrm{\COtwo}}P \\\nonumber &+RT\ln \varphi_{\textrm{\COtwo}}(T, P, y)
\end{align} 

where $\varphi$ is the fugacity coefficient, $\gamma$ is the activity coefficient, $y_\textrm{\COtwo}$ is the mole fraction of \COtwo in the vapor phase, and $P$, $T$, and $R$ have their usual meanings. They used the equation of state from \citet{Duan1992} to obtain the fugacity coefficient of pure \COtwo, ignoring the effect of water vapor. For Pitzer interaction parameters and standard chemical potential, they used parameterized relations.

\begin{table*}
\centering
\caption{Summary of modeling approaches for \COtwo solubility in water and brines}
\label{table:modeling}
\begin{threeparttable}
	\begin{tabular}{ccccc}

		\hline
	\textbf{Study} & \textbf{System} & \textbf{P-T range} & \textbf{Approach} \\\hline
	\cite{Li1986} & \makecell{Pure water\\NaCl brines} & \makecell{323.15--523.15K\\ 2--100MPa} & \makecell[l]{PR EOS, Henry's law,\\and SPT} \\\hline
	\cite{Harvey1989} & \makecell{Pure water\\Multi-salt brines} & \makecell{273.15--423.15K\\ 0--150MPa} & \makecell[l]{MSA, Born's equation} \\\hline
	\cite{Zuo1991} & \makecell{Pure water\\NaCl brines} & \makecell{273.15--533.15K\\ 0--150MPa} & \makecell[l]{PT EOS, Debye-Hückel\\electrostatic term} \\\hline
	\cite{Sorensen2002} & \makecell{Pure water\\NaCl, KCl, \CaCltwo brines} & \makecell{298.15--523.15K\\ 0.1--140MPa} & \makecell[l]{Cubic EOS, hypothetical\\critical properties for salts}  \\\hline
	\cite{Duan2003} & \makecell{Pure water\\NaCl brines} & \makecell{273--533K\\ 0--200MPa} & \makecell[l]{EOS from \citet{Duan1992}\\Pitzer activity model} \\\hline
	\cite{Duan2006} & \makecell{Pure water\\Multi-salt brines} & \makecell{273--533K\\ 0--200MPa} & \makecell[l]{EOS from \citet{Duan1992}\\Pitzer activity model} \\
		\hline
\end{tabular}
\end{threeparttable}

\end{table*}

\cite{Duan2006} presents an improved version over their original model. It introduces a non-iterative equation for calculation of \COtwo fugacity coefficients and also fitted the parameters to new experimental data. The new model was also expanded to multi-salt solutions. \COtwo solubility is then given by Equation \ref{eq:duan_cotwo_solubility}. Both \citet{Duan2003} and \citet{Duan2006} models show very good agreement with experimental data.

\begin{align}
\label{eq:duan_cotwo_solubility}\ln m_{\mathrm{CO}_2} & =  \ln y_{\mathrm{CO}_2} \varphi_{\mathrm{CO}_2} P-\mu_{\mathrm{CO}}^{1(0)} / R T \\
& -2 \lambda_{\mathrm{CO}_2-\mathrm{Na}}\left(m_{\mathrm{Na}}+m_{\mathrm{K}}+2 m_{\mathrm{Ca}}+2 m_{\mathrm{Mg}}\right) \nonumber\\
& -\zeta_{\mathrm{CO}_2-\mathrm{Na}-\mathrm{Cl}} m_{\mathrm{Cl}}\left(m_{\mathrm{Na}}+m_{\mathrm{K}}+m_{\mathrm{Mg}}+m_{\mathrm{Ca}}\right) \nonumber\\
& +0.07 m_{\mathrm{SO}_4}\nonumber
\end{align}

\subsection{Simulation approaches}
Beyond physical models, there exist more sophisticated simulation approaches for modeling chemical equilibrium and estimating \COtwo solubility. While fundamentally based on the same physical principles, simulation approaches often employ numerical methods and leverage computer algorithms to solve a complex system of equations that describe the system of interest. Simulation software are also more sophisticated and can simulate more complex scenarios, like equilibrium in the presence of other gases (e.g., impurities), or kinetic modeling. There are two main approaches for computer simulation of chemical equilibria: 1) Gibbs energy minimization (GEM), and 2) laws of mass-action (LMA) \citep{leal2017}.

In GEM approach, the goal is to identify the different species and their concentrations that minimize the system's Gibbs free energy. The problem can be expressed mathemtically as:

\begin{equation}
\min _n G=n^{\mathrm{T}} \mu \equiv \sum_{i=1}^{\mathrm{N}} n_i \mu_i \text { subject to }\left\{\begin{array}{l}
A n=b \\
n \geq 0
\end{array}\right.
\end{equation}

where $G$ is Gibbs free energy of the system, $\mu = [\mu_1, \ldots, \mu_m]^T$ is the vector of chemical potentials of the $N$ species.

Among the most popular software that uses GEM approach are THERIAC, ChemSage, HCh, FactSage, and GEM-Selektor. Software that uses LMA approach include WATEQ, MINEQL, WATEQ4F, MINTEQA2, EQ3/6, MINEQL+, CHESS, PHREEQC, The  Geochemist's Workbench, CHIM-XPT, and Selektor-C \citep{leal2017}.

However, in geochemical literature and software, the most common approach is to use LMA. In LMA, the equilibrium constant of a reaction is expressed as a function of the concentrations of the reactants and products. The equilibrium constant is then used to calculate the concentrations of the species at equilibrium. The problem can be expressed mathematically as:

\begin{equation}
	K_m = \prod^N_{i=1}a_i^{\nu_{mi}}
\end{equation}

where $K_m(P,T)$ is the equilibrium constant at pressure $P$ and temperature $T$, $a_i$ is the $i$th chemical species, $\nu_{mi}$ is the stoichiometric coefficient of the $i$th species in the $m$th reaction, and $N$ is the number of species \citep{leal2017}.

Simulation software that use LMA approach include WATEQ, MINEQL, WATEQ4F, MINTEQA2, EQ3/6, MINEQL+, CHESS, The Geochemist's Workbench, CHILLER, CHIM-XPT, SOLVEQ-XPT, PHREEQC, MIN3P, HYDROGEOCHEM, TOUGHREACT, CrunchFlow, and PFLOTRAN \citep{leal2017}.
\section{Factors controlling \COtwo solubility in water}
This section reviews the main factors that control the solubility of \COtwo in water: pressure, temperature, water chemistry, and also the presence of impurities. We discuss the mechanisms by which these factors affect solubility and to what extent.

\subsection{Effect of pressure and temperature}
Pressure and temperature are key factors governing the solubility of substances in solvents. In the case of \COtwo solubility in water, solubility increases with increasing pressure. Pressure effect is more pronounced at lower pressures, and becomes less sensitive at higher pressures when \COtwo is highly compressed. 

\begin{figure}[H]
    \centering
    \includegraphics[width=0.9\linewidth]{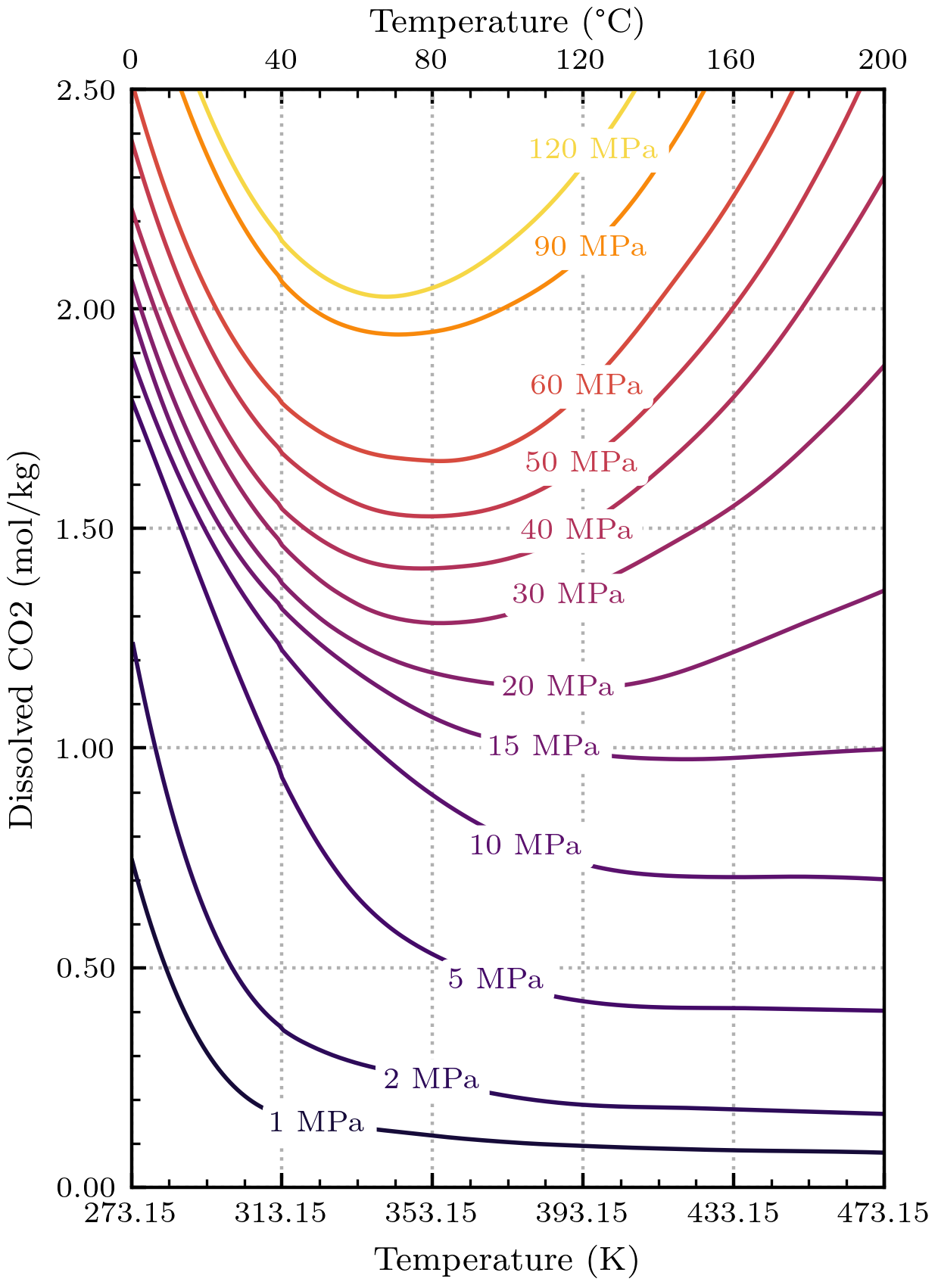}
    \caption{Isobaric solubilities of \COtwo in pure water with temperature.}
    \label{fig:solubility_with_pressure_and_temperature}
\end{figure}

\begin{figure}[H]
    \centering
    \includegraphics[width=0.9\linewidth]{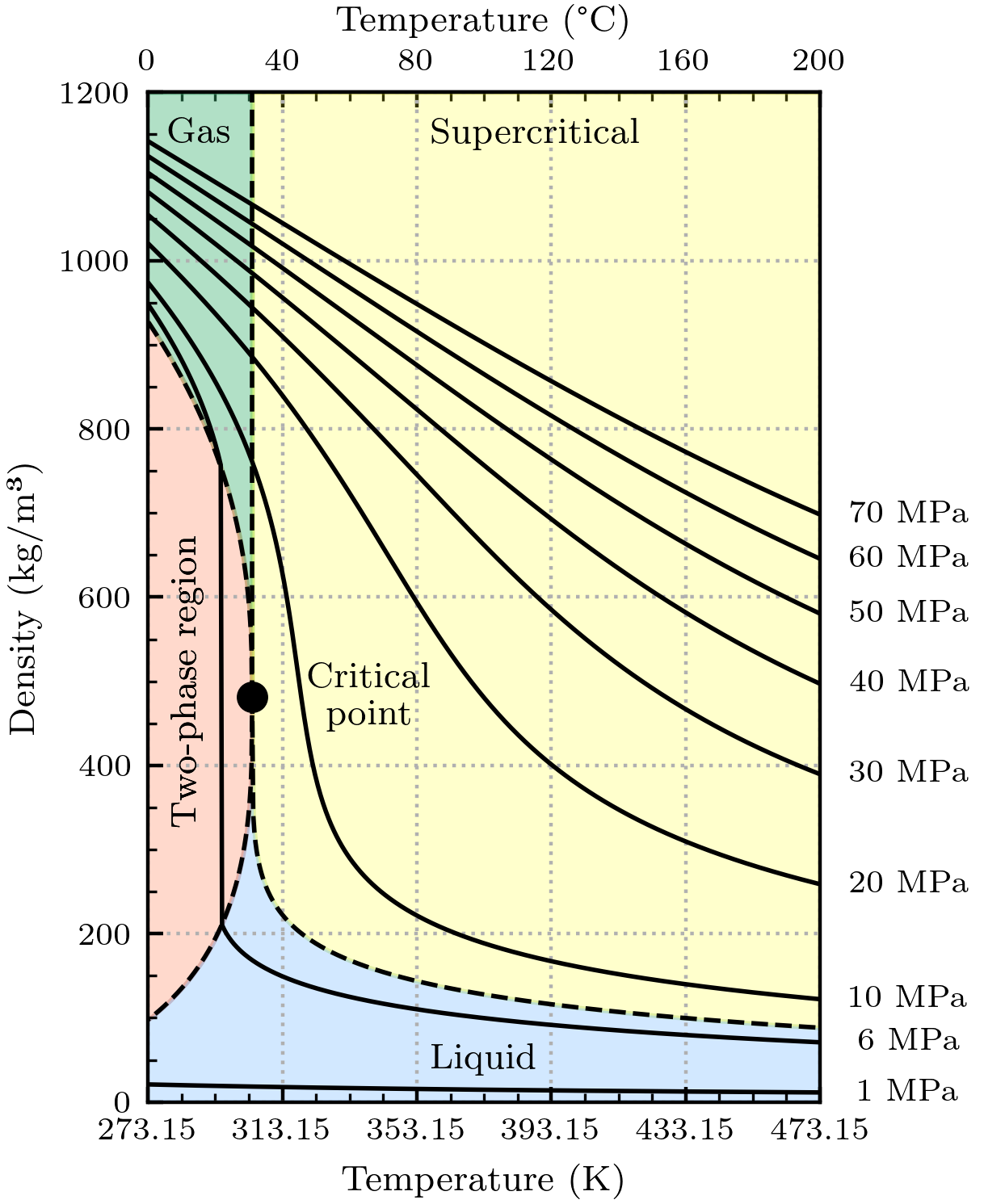}
    \caption{\COtwo density change with  changes in the temperature.}
    \label{fig:co2_density}
\end{figure}

At subcritical conditions ($P<7.38MPa$ and $T<304.2K$), solubility decreases with increasing temperature (Figure \ref{fig:solubility_with_pressure_and_temperature}). This is because higher temperatures reduce the gas's tendency to dissolve, as kinetic energy overcomes intermolecular attractions between water molecules and \COtwo. At supercritical conditions ($P\geq7.38MPa$ and $T\geq304.2K$), \COtwo density decreases sharply with increase in temperature (Figure \ref{fig:co2_density}) and solubility continues to decreases with temperature increase. At higher pressures, the density is less sensitive to temperature changes, the trend reverses and solubility starts increasing once more.

The effect of both pressure and temperature can be understood through the lens of the Gibbs free energy $\Delta G$ and favorability of reaction. Gibbs free energy is given by Equation \ref{eqn:gibbs_free_energy}:

\begin{equation}
    \Delta G = \Delta H - T\Delta S
    \label{eqn:gibbs_free_energy}
\end{equation}

A reaction will be spontaneous or favorable if $\Delta G < 0$, and unfavorable if $\Delta G > 0$. Figure \ref{fig:solubility_with_pressure_and_temperature} shows isobaric solubilities of \COtwo in pure water with temperature. Data was collected from \citet{Guo2014, Koschel2006, LaraCruz2020, Lucile2012, Messabeb2017, Spycher2003, Wang2019, Yan2011, Zhao2015, Zhao2015b, Zhao2015c, Dodds1956}, combined, plotted, interpolated, and smoothed to remove noise and better showcase the trends of \COtwo solubility. 

\subsection{Effect of water chemistry}
The solubility of \COtwo is sensitive to water composition, particularly the types and concentrations of dissolved aqueous species. While the effects of pressure and temperature on solubility are relatively straightforward, the influence of water chemistry is more nuanced. This section explores how different ions commonly found in natural formation brines alter \COtwo solubility.

The dissolution of \COtwo in water initiates a series of reversible reactions known as the carbonate system. First, gaseous \COtwo dissolves to become an aqueous species, \COtwoaq (Equation~\ref{eq:co2_gas_to_aqueous}). This aqueous \COtwo then reacts with water to form carbonic acid, \carbonic (Equation~\ref{eq:carbonic_acid_formation}). Carbonic acid, being a weak acid, subsequently dissociates into a proton, \proton, and a bicarbonate ion, \bicarbonate (Equation~\ref{eq:bicarbonate_formation}). Finally, bicarbonate can dissociate further into another proton and a carbonate ion, \carbonate (Equation~\ref{eq:carbonate_formation}).

\begin{align}
    CO_2(g) &\rightleftharpoons CO_2(aq) \label{eq:co2_gas_to_aqueous} \\
    CO_2(aq) + H_2O(l) &\rightleftharpoons H_2CO_3(aq) \label{eq:carbonic_acid_formation} \\
    H_2CO_3(aq) &\rightleftharpoons HCO_3^-(aq) + H^+(aq) \label{eq:bicarbonate_formation} \\
    HCO_3^-(aq) &\rightleftharpoons CO_3^{2-}(aq) + H^+(aq) \label{eq:carbonate_formation}
\end{align}

\begin{figure}[H]
    \includegraphics[width=\linewidth]{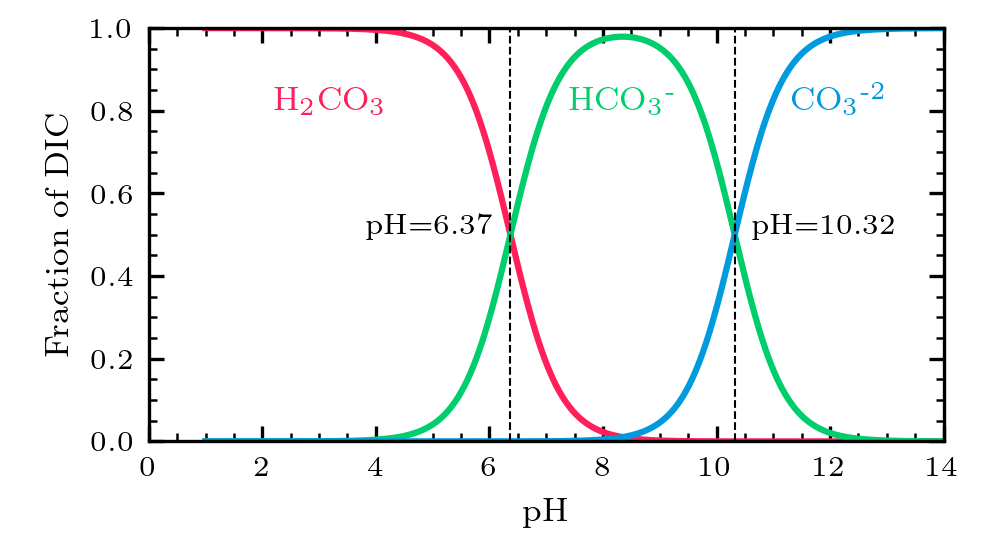}
    \caption{Distribution of carbon species in water as a function of pH (at $T=25\degree C$)}
    \label{fig:carbonate}
\end{figure}

The relative abundance of these carbon species is a function of the solution's pH, as illustrated in Figure~\ref{fig:carbonate}. Under typical geological storage conditions, the brine is acidic (pH < 4.0), and the vast majority of dissolved carbon exists as the neutral aqueous species, \COtwoaq \citep{DeSilva2015, Gilbert2016}. It is this neutral species that is most directly affected by the presence of dissolved salts.

\begin{figure}[H]
    \centering
    \includegraphics[width=\linewidth]{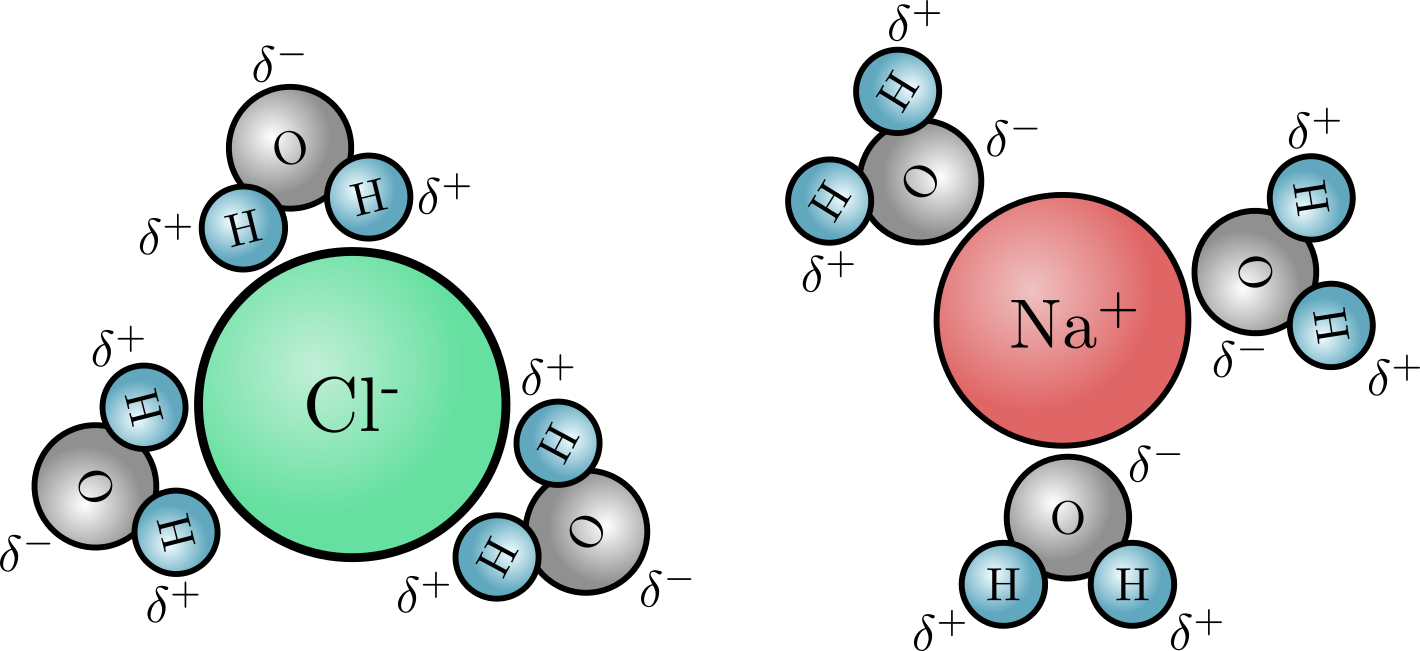}
    \caption{Schematic of hydration shells around ions.}
    \label{fig:hydration_shell}
\end{figure}

The presence of dissolved salts reduces the solubility of gases like \COtwo. This phenomenon is known as the \textit{salting-out} effect. It occurs because the dissolved ions attract polar water molecules, forming stable structures called \textit{hydration} or \textit{solvation shells} around themselves (Figure~\ref{fig:hydration_shell}). This process, sometimes referred to as electrostriction, effectively "locks up" water molecules, reducing the amount of free solvent available to dissolve other substances like \COtwo \citep{Liu2011}.

The strength and structure of these hydration shells depend on the specific properties of the ion, primarily its charge and size (ionic radius). Ions with a higher charge density (higher charge and/or smaller radius) exert a stronger electrostatic pull on water molecules, leading to a more pronounced salting-out effect \citep{Hribar2002, Gorgenyi2006}. For example \sodium, with its smaller ionic radius, has a higher charge density than potassium and thus creates a more ordered hydration shell, resulting in a stronger salting-out effect. Similarly, \magnesium has a smaller ionic radius (72 pm) than \calcium (106 pm) \citep{Shannon1976}. Despite both having a +2 charge, the higher charge density of \magnesium leads to slightly stronger hydration \citep{Friesen2019}.

A common approach to quantify the overall concentration of salts in a brine is to calculate its molal ionic strength, $I$, given by Equation~\ref{eq:ionic_strength}, where $b_i$ is the molality of ion $i$ and $z_i$ is its charge.

\begin{equation}
    I = \frac{1}{2} \sum_i^n b_iz_i^2
    \label{eq:ionic_strength}
\end{equation}

Many studies have simplified experimental work by assuming that ionic strength is the main factor controlling the salting-out effect. This implies that any brine can be represented by a simple NaCl solution of equivalent ionic strength \citep{Portier2005, Zhao2015c}. However, this is often an oversimplification.

\begin{figure}[H]
    \includegraphics[width=\linewidth]{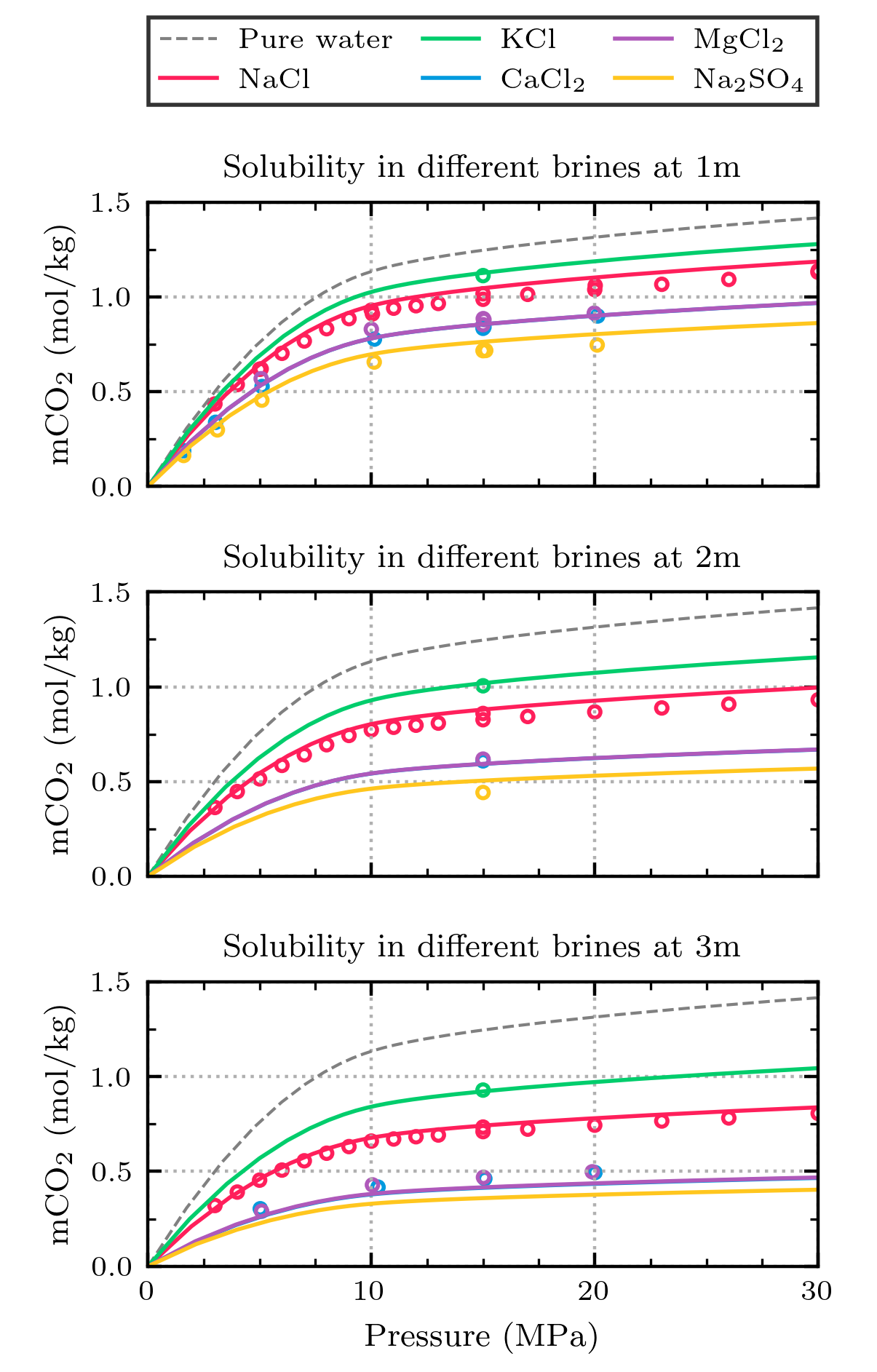}
    \caption{\COtwo solubility in different solutions with concentrations from 1m to 3m at 323.15K}
    \label{fig:solubility_1_3_molal}
\end{figure}

Figure~\ref{fig:solubility_1_3_molal} demonstrates that solutions with the same ionic strength can exhibit significantly different salting-out effects. Lines are simulation data obtained from PHREEQC using \texttt{pitzer.dat}, while data points are experimental measurements from \citet{Wang2019, Zhao2015b, dosSantos2020, dosSantos2021, LaraCruz2020, Messabeb2017}. For instance, NaCl and KCl solutions have identical ionic strengths at a given molality, as do \MgCltwo and \CaCltwo solutions, yet their impact on \COtwo solubility varies. The experimental data show that the salting-out strength for these salts increases in the order: KCl < NaCl < \MgCltwo < \CaCltwo < \NaSOfour. This proves that ionic strength alone is not a sufficient predictor and that the specific nature of the ions involved is crucial.

Since the physical mechanism of salting-out is tied to water molecules being bound in hydration shells, a more direct metric should provide better predictions. \citet{Gilbert2016} proposed using the molality of \textit{electrostricted water}, $h_a$, which is the total concentration of water molecules engaged in hydrating ions. It is calculated using Equation~\ref{eqn:electrostricted}, where $C_i$ is the concentration of ion $i$ and $h_{comp}$ is its characteristic hydration number.

\begin{equation}
    h_a = \sum C_in_ih_{comp}
    \label{eqn:electrostricted}
\end{equation}

\begin{figure}[H]
    \centering
    \includegraphics[width=0.9\linewidth]{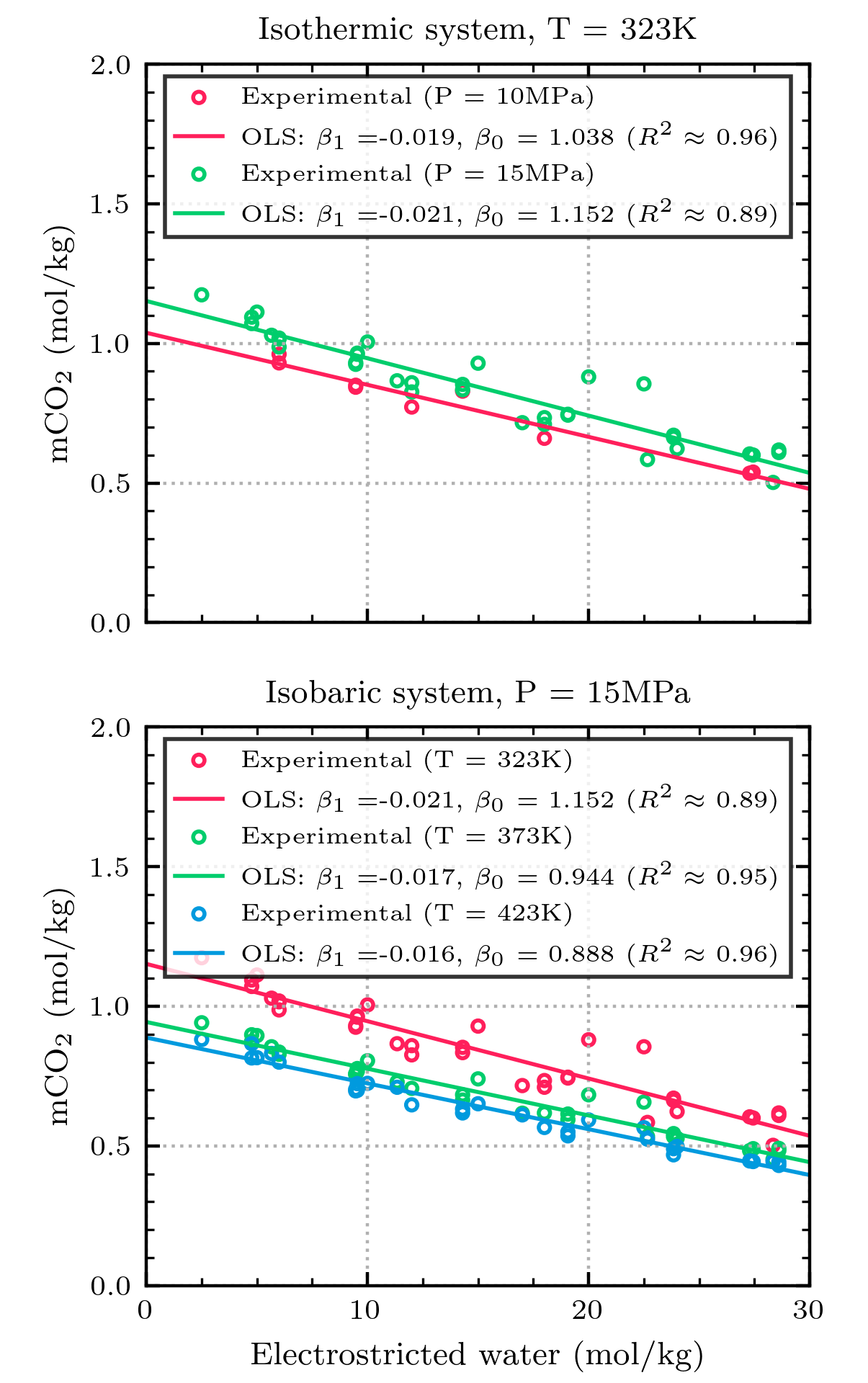}
    \caption{Isothermal (top) and isobaric (bottom) solubilities of \COtwo as a function of molality of electrostricted water.}
    \label{fig:electrostriction}
\end{figure}

As shown in Figure~\ref{fig:electrostriction}, \COtwo solubility exhibits a strong linear correlation with the molality of electrostricted water. The high coefficients of determination ($R^2 \geq 0.89$) across a wide range of experimental data confirm that $h_a$ is a much stronger predictor of the salting-out effect than ionic strength. Experimental data for the visualization is from \citep{dosSantos2020,dosSantos2021,Kamps2007,Koschel2006,LaraCruz2020,Liu2011,Messabeb2017,Rumpf1993,Tong2013,Wang2019,Yan2011,Zhao2015,Zhao2015b,Zhao2015c, Bermejo2005}.

Despite its predictive power, this approach has limitations. It assumes that the hydration effect of each ion is simply additive, whereas studies have shown that ion hydration can be a cooperative and non-additive process \citep{Tielrooij2010}. Furthermore, determining accurate and consistent hydration numbers ($h_{comp}$) is challenging, with values often varying between studies \citep{Marcus2014}.

Finally, at higher salt concentrations, another factor comes into play: the formation of \textit{ion pairs} \citep{Ohtaki1993}. Ions may associate to form neutral pairs (e.g., NaCl\textsuperscript{0}) or charged complexes (e.g., MgCl\textsuperscript{+}, CaCl\textsuperscript{+}, NaSO\textsubscript{4}\textsuperscript{-}). These pairs have lower effective charge densities than free ions, altering their interaction with water and thus adding another layer of complexity to the salting-out phenomenon \citep{Chen2015, Gilbert2016}.

\subsection{Effect of impurities in \COtwo streams}
The presence of impurities in a captured \COtwo stream is a critical factor influencing the feasibility, safety, and efficiency of geological storage projects. The composition of the \COtwo stream, which varies depending on the capture technology, directly impacts the physical and chemical properties of the injected fluid (Table~\ref{table:expected-co2-stream}). These changes, in turn, alter the storage capacity of the geological formation \citet{razak2023}. The primary effects can be categorized based on the type of impurity, with non-condensable and condensable gases exhibiting distinct behaviors.

\begin{table}[H]
\caption{Expected \COtwo stream composition \citep{Frontier2023}}
\label{table:expected-co2-stream}
\begin{tabular}{ccc}
	\hline
Component & ppmv & mol\%  \\
	\hline
    \COtwo & 983,960  & 98.3960 \\
    Water (H\textsubscript{2}O) & 14,389 & 1.4389 \\
    Nitrogen (N\textsubscript{2}) & 855 & 0.0855 \\
    Oxygen (O\textsubscript{2}) & 767 & 0.0767 \\
    Total hydrocarbons (as CH\textsubscript{4}) & 12 & 0.0012 \\
    Total Sulfur (as S) & 11 & 0.0011 \\
    Hydrogen (H\textsubscript{2}) & 6 & 0.0006 \\\hline

\end{tabular}
\end{table}

\subsubsection{Non-condensable gases}
Common non-condensable impurities, such as nitrogen (\Ntwo), oxygen (\Otwo), and argon (Ar), are frequently found in \COtwo streams, particularly from oxy-fuel combustion processes. These impurities significantly reduce the storage capacity of a reservoir compared to the injection of pure \COtwo \citep{wang2015Effect}. The primary mechanism for this reduction is a decrease in the density of the \COtwo stream. Because non-condensable gases do not compress to the same degree as pure \COtwo under typical storage pressures and temperatures, they occupy a larger volume for the same mass, thus lowering the overall density of the mixture \citep{razak2023, IEAGHG2011}.
This reduction in storage capacity can be substantial. Studies indicate that a \COtwo stream containing approximately 15\% non-condensable impurities can reduce the storage capacity by as much as 40\% in shallow reservoirs compared to pure \COtwo \citep{IEAGHG2011}. The magnitude of this effect is highly dependent on the pressure and temperature of the reservoir. The capacity reduction is most pronounced in shallow, low-pressure formations and diminishes at greater depths where higher pressures increase the compressibility of the entire gas mixture \citep{razak2023, IEAGHG2011}.

\subsubsection{Condensable gases}
In contrast to non-condensable gases, certain condensable impurities can have a different effect on storage capacity. Sulphur dioxide (\SOtwo), a common impurity, is more easily condensable than \COtwo and has been shown to increase the density of the \COtwo mixture. This can lead to a slight increase in the mass of \COtwo that can be stored in a given volume \citep{wang2015Effect}. Modeling has suggested that under certain pressure conditions, the presence of \SOtwo can create more storage space for \COtwo, an effect attributed to its ability to decrease the average distance between the molecules in the mixture \citep{IEAGHG2011}.

However, the potential physical benefit of \SOtwo is often outweighed by its significant chemical drawbacks. \SOtwo readily dissolves in water to form highly corrosive acids, primarily sulphuric acid. This can lead to severe corrosion of well materials and can react with reservoir rock to dissolve existing minerals and precipitate new ones, such as anhydrite or gypsum \citep{xiang2018}. This precipitation can block pore throats, reducing the reservoir's permeability and injectivity, which negatively impacts the overall storage potential and long-term security \citep{razak2023, IEAGHG2011}.

Hydrogen sulfide (\HtwoS), another common impurity, has a more complex effect. While it is more condensable than non-condensable gases, it generally leads to a decrease in storage capacity, though less severe than that caused by gases like \Ntwo or Ar \citep{IEAGHG2011}. Similar to \SOtwo, \HtwoS also poses a significant corrosion risk to pipelines and well infrastructure \citep{xiang2018}.

In summary, the effect of impurities on \COtwo storage capacity is highly dependent on their physical properties. Non-condensable gases like \Ntwo, \Otwo, and Ar decrease capacity by lowering the stream's density, whereas a condensable gas like \SOtwo can physically increase it. However, the chemical reactivity of impurities like \SOtwo and \HtwoS introduces critical risks of corrosion and adverse rock-fluid interactions that must be carefully managed to ensure the integrity and safety of the storage site.

\section{Improved Pitzer interaction parameters for \COtwo solubility}
\begin{figure}[H]
	\centering
	\includegraphics[width=\linewidth]{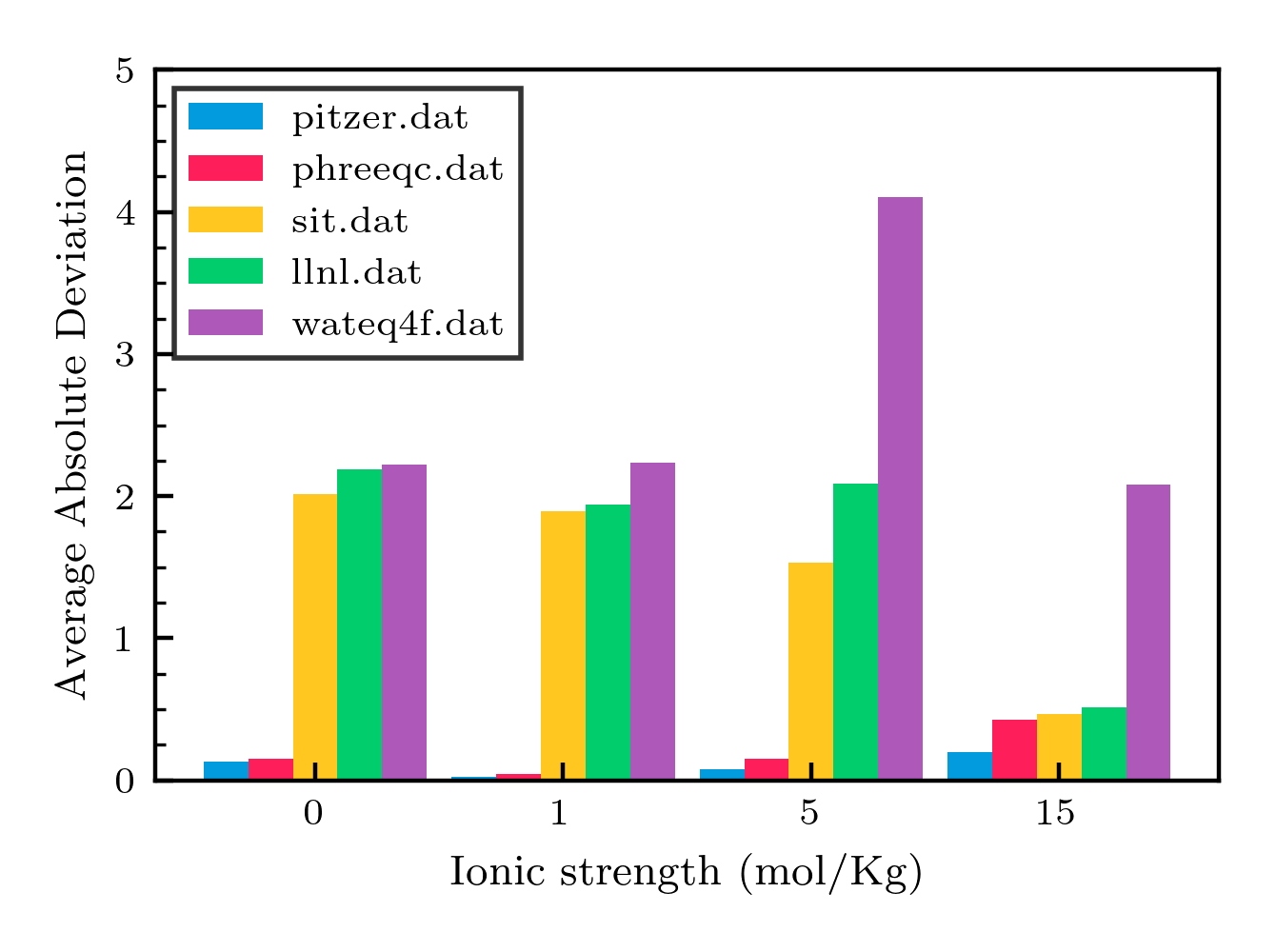}
	\caption{Error of PHREEQC databases at different ionic strengths}
	\label{fig:phreeqc_databases_benchmark_w_ionic_strength}
\end{figure}

PHREEQC offers a suite of thermodynamic databases for equilibrium and kinetic simulations of hydrogeochemical reactions. These databases differ primarily in their validity within different pressure and temperature ranges, support for different aqueous species, and the fugacity and activity models employed to account for the non-ideality of aqueous and gaseous phases. \citet{lu2022} provided a more comprehensive comparison of these different thermodynamic databases. Some of the most commonly used databases include \texttt{pitzer.dat}, \texttt{phreeqc.dat}, \texttt{llnl.dat}, \texttt{sit.dat}, and \texttt{wateq4f.dat}. We evaluated the different databases on the compiled experimental data for \COtwo solubility in brines. Figure \ref{fig:phreeqc_databases_benchmark_w_ionic_strength} illustrates the error of various PHREEQC databases benchmarked against experimental data at different ionic strengths.

Figure \ref{fig:phreeqc_databases_benchmark_w_ionic_strength} clearly demonstrates that both \texttt{pitzer.dat} and \texttt{phreeqc.dat} exhibit the lowest error in comparison to other databases. Notably, \texttt{pitzer.dat} consistently maintains a low error even at higher ionic strengths, suggesting a more robust activity model. This stems from the fact that \texttt{pitzer.dat} employs the more sophisticated Pitzer (as the name suggests) activity model, while \texttt{phreeqc.dat} utilizes a mixed WATEQ and Davies equation. 

%\begin{table}[H]
%    \caption{Benchmark of different databases included in PHREEQC}
%	\label{table:phreeqc_databases_benchmark}
%    \centering
%\begin{tabular}{cc}
%    \hline
%    Database & Absolute Average Deviation\\
%    \hline
%	\texttt{phreeqc.dat} & 0.11978\\
%   \texttt{pitzer.dat} & 0.16160 \\
%	\texttt{llnl.dat} & 1.33562 \\
%	\texttt{sit.dat} & 1.71063 \\
%	\texttt{wateq4f.dat} & 1.26929 \\
%    \hline
%\end{tabular}
%\end{table}

Within Pitzer model, the activity coefficient of dissolved \COtwo is governed by the binary interaction parameter ($\lambda$) between neutral species (i.e., \COtwo) and individual cations/anions, and by the ternary interaction parameters ($\zeta$) between neutral species and cations and anions. The activity coefficient of \COtwo is given by Equation \ref{eq:cotwo_activity}.

\begin{align}
\label{eq:cotwo_activity}ln\gamma_{CO_2} = & \sum_c 2\lambda_{CO_2-c} m_c+\sum_a 2  \lambda_{CO_2-a} m_a \\ \nonumber &+ \sum_c\sum_a \zeta_{CO_2-a-c} m_c m_a
\end{align}

where $c$ and $a$ denote cations and anions, respectively. The $\lambda$ and $\zeta$ parameters are usually determined through fitting experimental data. While some authors consider these parameters to be temperature- and pressure-dependent \citep{Duan2003}, others focus solely the temperature dependence \citep{appelo2015}. In PHREEQC (combined with \texttt{pitzer.dat}), where only the temperature dependence is considered, the interaction parameters are given by the polynomial in Equation \ref{eqn:temperature_poly}.

\begin{align}
	\label{eqn:temperature_poly}& A_0 + A_1  \left(\frac{1}{T}–\frac{1}{T_R}\right) + A_2  \ln\left(\frac{T}{T_R}\right) \\\nonumber &  + A_3  (T–T_R)  + A_4  (T^2–T_R^2) + A_5  \left(\frac{1}{T^2}–\frac{1}{T_R^2}\right)
\end{align}

where $T_R = 298.15$ and $A_0$ to $A_5$ are fitting parameters obtained by optimizing against experimental data. While in PHREEQC+\texttt{pitzer.dat} the interaction parameters can be expressed as functions of temperature; the default $\lambda_{CO_2-a}$, $\lambda_{CO_2-c}$, and $\zeta_{CO_2-a-c}$ are constants and do not consider any temperature dependence.

When examining \texttt{pitzer.dat}, the first thing to notice is that the parameters $\lambda_{CO_2-Mg^{2+}}$ and $\lambda_{CO_2-Ca^{2+}}$ have the same value ($\lambda_{CO_2-Mg^{2+}}=\lambda_{CO_2-Ca^{2+}}=0.183$), However, experimental data suggests a small but discernible difference, with the salting-out of \COtwo in \MgCltwo solutions being a bit higher than that in \CaCltwo solutions at equal molality. Furthermore, a systematic deviation from experimental values of dissolved \COtwo in NaCl and \NaSOfour brines is observed at higher concentrations.

\begin{figure}[H]
	\centering
	\includegraphics[width=0.45\textwidth]{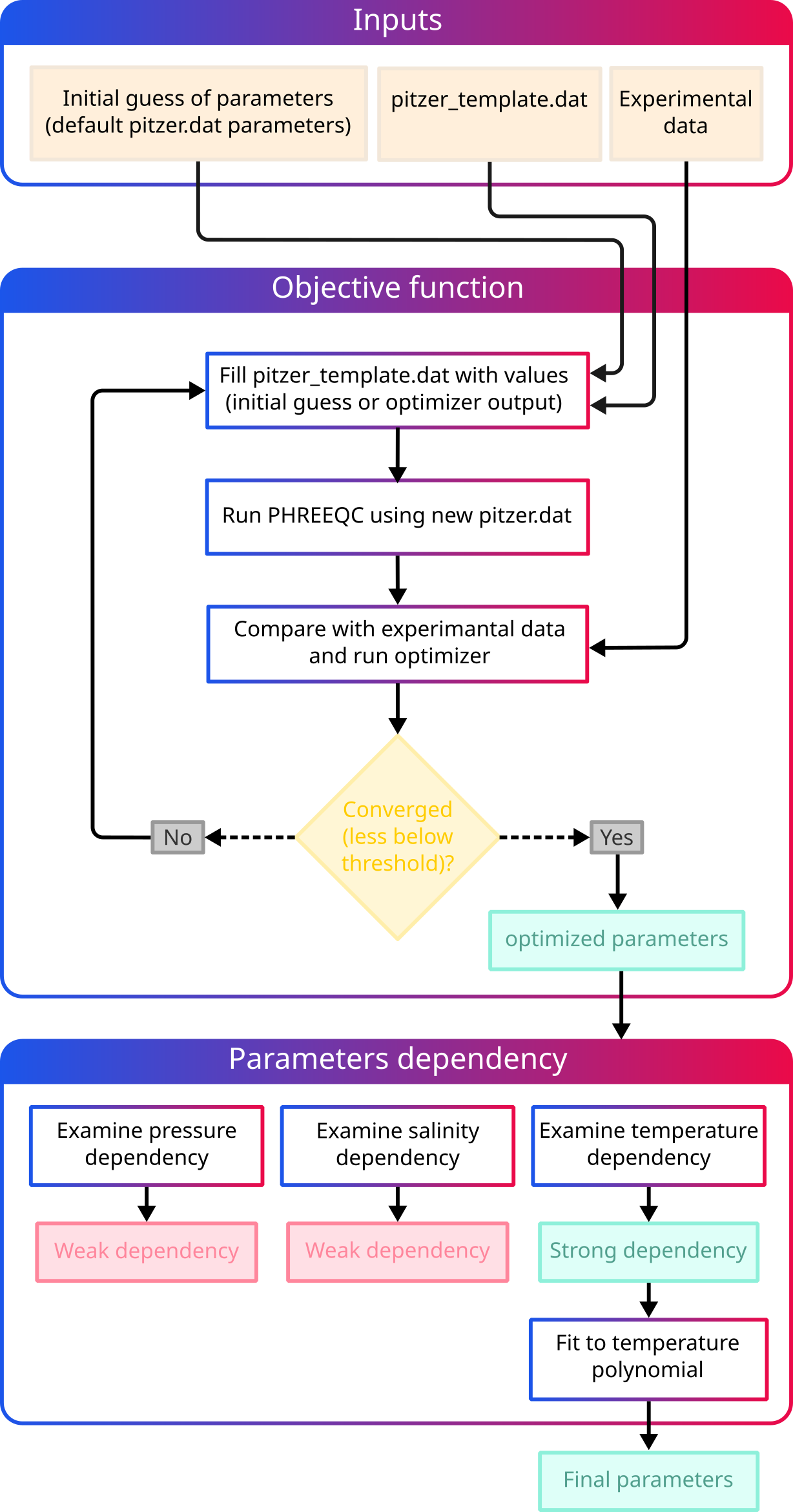}
	\caption{Workflow for obtaining new Pitzer parameters for \COtwo solubility in brines.}
	\label{fig:pitzer-parameters}
\end{figure}

To address these discrepancies, we leveraged recently published experimental data to obtain a new and improved set of Pitzer interaction parameters. Our workflow, illustrated in Figure \ref{fig:pitzer-parameters}, involves a Python script that iteratively modifies a \texttt{pitzer.dat} database template. The script then proceeds to run PHREEQC using the database with modified parameters, compares PHREEQC results to selected experimental data, and optimizes the parameters accordingly. The new parameters are again used to populate the database template and the procedure continues until convergence. For optimization, we used \texttt{optimize.minimize()} function from \texttt{scipy} library, where the objective function is the L2 norm of the residuals between model predictions and experimental data.

To quantify the improvement of the new parameters over the default ones, we used the percent change in Average Absolute Deviation (AAD\%) metric, which is defined as:

\begin{equation}
AAD^{\text{default}} = \frac{100}{N_p} \sum_{i=1}^{N_p} \frac{ \left| y_i^{\text{default}} - y_i^{\text{exp}} \right| }{ y_i^{\text{exp}} }
\end{equation}

\begin{equation}
AAD^{\text{mod}} = \frac{100}{N_p} \sum_{i=1}^{N_p} \frac{ \left| y_i^{\text{mod}} - y_i^{\text{exp}} \right| }{ y_i^{\text{exp}} }
\end{equation}

\begin{equation}
\label{eqn:aad_improvement}
\text{AAD\%} = \frac{ AAD^{\text{default}} - AAD^{\text{mod}} }{ AAD^{\text{default}} } \times 100
\end{equation}

Our optimization began by considering \HtwoO-\COtwo-NaCl system. The parameters governing the activity of \COtwo in this system are $\lambda_{CO_2-Na^+}$, $\lambda_{CO_2-Cl^-}$, and $\zeta_{CO_2-Na^+-Cl^-}$. In similar studies, the authors often set $\lambda_{CO_2-Cl^-}$ to 0, and proceed to optimize the other parameters \citep{dosSantos2021b}. In contrast, we retained the default value in PHREEQC  ($\lambda_{CO_2-Cl^-}=-0.05$). We then fitted the parameters $\lambda_{CO_2-Na^+}$ and $\zeta_{CO_2-Na^+-Cl^-}$ using data from \citep{Wang2019,LaraCruz2020, Rumpf1994}, which was selected for its consistency and quality. By default, \texttt{pitzer.dat} does not have a value for $\zeta_{CO_2-Na^+-Cl^-}$. We observed that omitting this parameter leads $\lambda_{CO_2-Na^+}$ to be strongly dependent on NaCl concentration, which is consistent with findings in similar studies \citep{dosSantos2021b}. Therefore, adding $\zeta_{CO_2-Na^+-Cl^-}$ was crucial to minimize parameters dependencies, allowing us to focus on temperature dependence. To prevent over-parametrization, we chose to fit only the first two terms ($A_0$ and $A_1$) in the polynomial in Equation \ref{eqn:temperature_poly} for both $\lambda_{CO_2-Na^+}$ and $\zeta_{CO_2-Na^+-Cl^-}$. The new parameters resulted in an AAD improvement of 75.92\% over the default values.

\begin{table*}[]
	\centering
	\caption{New Pitzer binary and ternary interaction parameters for $CO_2$ in different ionic systems. AAD Improvement \% quantifies the enhanced accuracy of the new parameters over the default values when benchmarked on experimental data.}
	\label{table:pitzer-new-parameters}
	\begin{threeparttable}

		\begin{tabular}{ccccccc}
			\hline
			System                     & Parameter                     & $A_0$       & $A_1$       & AAD improvement \%       \\
			\hline
			\multirow{2}{*}{NaCl}      & $\lambda_{CO_2-Na^+}$         & 0.11840622  & 11.42424414 & \multirow{2}{*}{75.92\%} \\\cline{2-4}
			                           & $\zeta_{CO_2-Na^+-Cl^-}$      & -0.01418353 & 3.01231593  &                          \\\hline
            \multirow{2}{*}{\NaSOfour} & $\lambda_{CO_2-SO_4^{-2}}$   & 0.08063083  & 60.42023486 & \multirow{2}{*}{71.55\%} \\\cline{2-4}
			                           & $\zeta_{CO_2-Na^+-SO_4^{-2}}$ & -0.0141691  & 10.56399699 &                          \\\hline
            \multirow{2}{*}{\KCl} & $\lambda_{CO_2-K^+}$   & 0.06554669  & 0 & \multirow{2}{*}{15.27\%} \\\cline{2-4}
			                           & $\zeta_{CO_2-K^+-Cl^-}$ & -0.00960956  & 0 &                          \\\hline
            \multirow{2}{*}{\CaCltwo}  & $\lambda_{CO_2-Ca^{2+}}$      & 0.18475566  & 21.97468005 & \multirow{2}{*}{16.14\%} \\\cline{2-4}
			                           & $\zeta_{CO_2-Ca^{2+}-Cl^-}$   & 0           & 0           &                          \\\hline
			\multirow{2}{*}{\MgCltwo}  & $\lambda_{CO_2-Mg^{2+}}$      & NA           & NA           & \multirow{2}{*}{NA}     \\\cline{2-4}
			                           & $\zeta_{CO_2-Mg^{2+}-Cl^-}$   & NA           & NA           &                          \\\hline
			
		\end{tabular}
	\end{threeparttable}
\end{table*}

For the \HtwoO--\COtwo--\NaSOfour system, we used the $\lambda_{CO_2-Na^+}$ parameter obtained from the previous step and proceeded to determine  $\lambda_{CO_2-SO_4^{-2}}$ and $\zeta_{CO_2-Na^+-SO_4^{-2}}$ using data from \citet{dosSantos2020,Rumpf1993,Zhao2015b}. The new parameters achieved an AAD\% improvement of 71.55\%.

For \HtwoO--\COtwo--\KCl system, we used data from \citet{Kamps2007,Zhao2015b,Liu2011}. The new parameters achieved an AAD\% improvement of 15.27\% over the default values when compared to experimental data.

For \HtwoO--\COtwo--\CaCltwo system, we used data from \citep{Tong2013,Zhao2015b}. The new parameters achieved improvement even when evaluated on data from \citep{Liu2011,LaraCruz2020, Messabeb2017} that was excluded from the fitting process because its inconsistency with the other experimental studies. The new parameters achieved an AAD improvement of 16.14\% over the default values when compared to experimental data.

Finally, for \HtwoO--\COtwo--\MgCltwo system, insufficiency of experimental data prevented achieving any meaningful or statistically significant improvement over the default values.

Our results show significant improvement for systems with abundant experimental data (e.g., \HtwoO--\COtwo--NaCl and \HtwoO--\COtwo--\NaSOfour). This emphasizes the importance of high quality lab data and the need for more studies related to the less studied systems like \HtwoO--\COtwo--\MgCltwo. The new parameters, and the AAD\% improvement over default ones, are shown in Table \ref{table:pitzer-new-parameters}.

To assess the robustness of the new parameters, we evaluated them on data from multi-salt solutions. This data was not used for fitting the parameters, and was only used for evaluation. In NaCl + KCl system, the new parameters achieved an AAD Improvement of 21.02\% (data from \cite{Liu2011,Tong2013}), while for NaCl + KCl + \CaCltwo, the AAD improvement was 15.09\% (data from \cite{Liu2011,Zhao2015c}). This finding is significant, as it demonstrates the applicability and predictive capability of the new parameters for \COtwo solubility in complex multi-salt solutions, a common scenario in subsurface environments.

\section{Physics-informed machine learning modeling}

Physics-informed machine learning refers to the integration of physical laws into machine learning models through several means \citep{Latrach2024}. This integration ensures abidance by known physical laws and may even allow the model to reliably extrapoalte beyond its training data. In this study, we designed a physics-informed machine learning model for \COtwo solubility prediction. The motivation is to produce a model that has a lower error than state-of-the-art models in the literature by directly integrating experimental data into the training process of a model whose architecture encodes the physical process of \COtwo dissolution. The subsequent sections detail the model architecture, data generation, training, and evaluation.

\subsection{Model architecture}
\begin{figure}[H]
	\centering
	\includegraphics[width=0.9\linewidth]{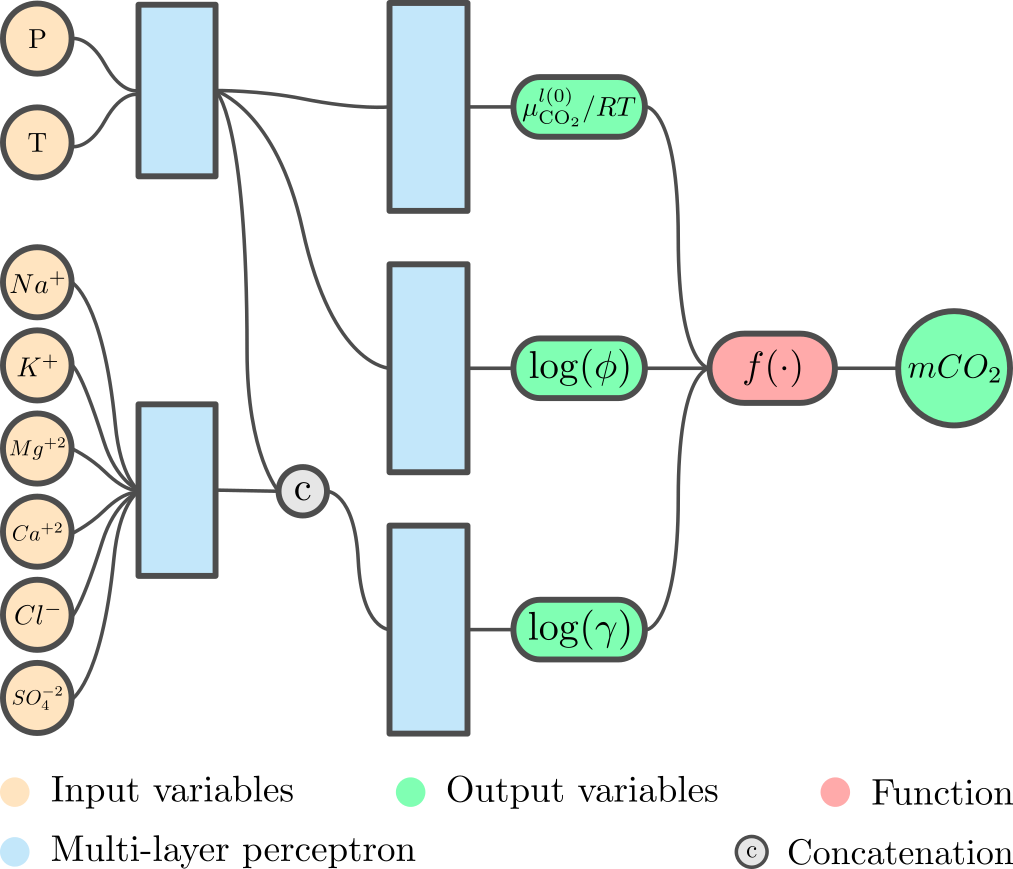}
	\caption{Architecture of multitask machine learning model for predicting \COtwo solubility in aqueous solutions.}
	\label{fig:ml-architecture}
\end{figure}

The machine learning model is based on a multi-tasking approach that integrates physical knowledge of the system into the model's architecture. Based on the work of \citet{Duan2003}, we envisioned a model that has two trunk networks: 1) a trunk that takes pressure and temperature inputs, and 2) a trunk that takes concentrations (i.e., molalities) of ionic species as inputs. The reasoning behind this two-trunks architecture is that our intermediary outputs (i.e., log of fugacity coefficient and chemical potential) are functions of pressure and temperature only, and using ionic concentrations as an input would be inappropriate and can cause the model to learn spurious correlations. The pressure--temperature trunk is then fed into two branch networks that separately output the log of the fugacity coefficient, $\log(\phi$), and the standard chemical potential (divided by $RT$), $\mu_{\textrm{\COtwo}}^{l(0)}/RT$. The outputs from the two trunks are concatenated and fed into a third branch that outputs the log of the activity coefficient $\log(\gamma)$ of \COtwoaq. Concatentation assures that the activity branch learns any temperature and pressure dependencies of the activity coefficient. All these branch and trunk networks are multilayer percepterons. For the sake of this discussion, we will call chemical potential, fugacity and activity coefficients as intermediary variables, and dissolved \COtwo as output variable. Figure~\ref{fig:ml-architecture} and Table~\ref{table:model-architecture} illustrate and summarize the model's architecture.

To establish a baseline to benchmark our proposed architecture, we also trained a simple fully-connected network, which we call \textit{blackbox model}. This is to ensure that our proposed architecture actually has a benefit over a simple neural network.

\begin{table*}[]

    \centering
    \caption{Summary of the multitask‐learning network: a shared P–T trunk feeding two thermodynamic heads ($\gamma$, $\mu$) and an activity branch (using both P–T and concentration features).}
    \begin{tabular}{l|cc|ccc}
        \hline
        Parameter           & P--T Trunk    & C Trunk       & $\log(\phi)$ branch & $\mu_{\textrm{\COtwo}\textsuperscript{l(0)}}/ RT$ branch & $\log(\gamma)$ branch \\
        \hline
        Input shape         & (batch, 2)    & (batch, 6)    & (batch, 64)     & (batch, 64)                & (batch, 96)    \\
        Output shape        & (batch, 64)   & (batch, 32)   & (batch, 1)      & (batch, 1)                 & (batch, 1)     \\
        \#Linear layers     & 2             & 2             & 3               & 3                          & 3              \\
        Hidden layers       & 2             & 2             & 2               & 2                          & 2              \\
        Neurons per layer   & [128, 64]     & [64, 32]      & [32, 16]        & [32, 16]                   & [64, 32]       \\
        Activation          & ReLU          & ReLU          & ReLU            & ReLU                       & ReLU           \\
        \hline
    \end{tabular}
	\label{table:model-architecture}
\end{table*}

\subsection{Data generation and training}
\begin{figure}[H]
	\centering
	\includegraphics[width=\linewidth]{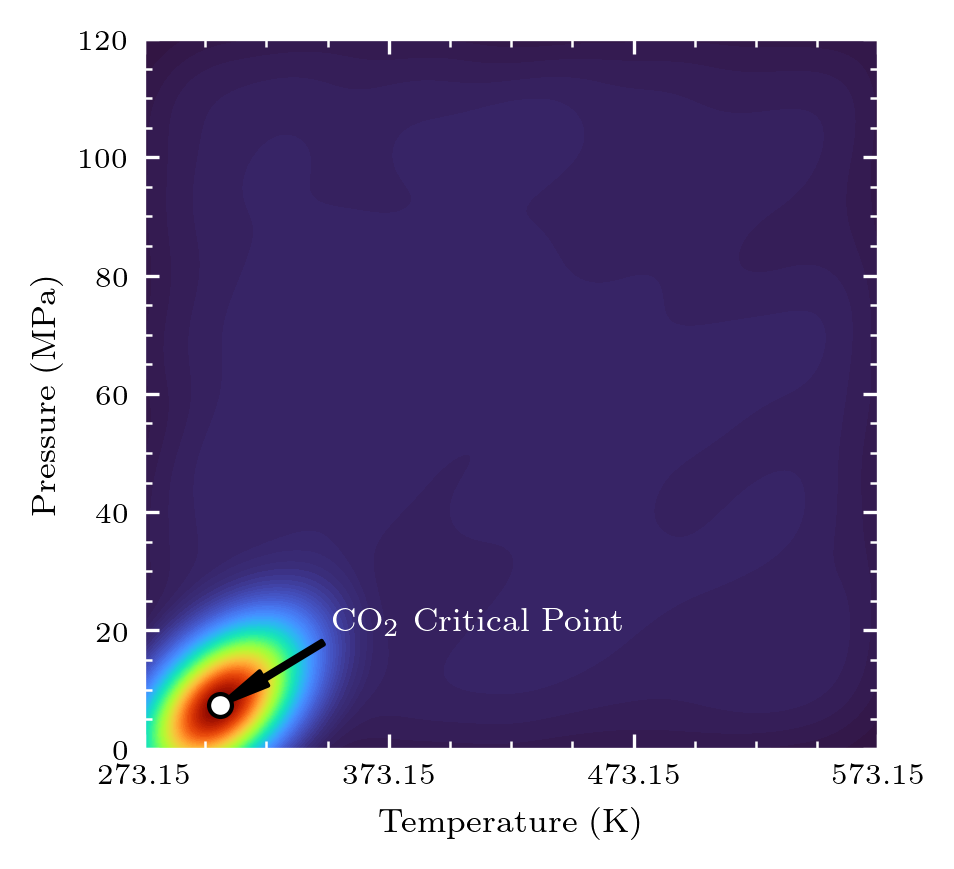}
	\caption{Density of generated data points for the synthetic training dataset.}
	\label{fig:data-generation}
\end{figure}

The first step into training this model involved synthetic data generation using the \citet{Duan2006} model. The model was used to generate a large dataset of intermediary variables as well as the output variable at a wide range of pressures, temperatures, and salinities. Due to neural networks' tendency to underfit sharp transitions in the target function, we followed a non-linear samplign strategy to generate more data points around the point of phase change from sub- to supercritical. The generated data for pressure and temperature has a Gaussian bump centerd around the critical point, with uniform sampling away from it. Figure~\ref{fig:data-generation} shows the joint distribution of the generated samples' pressures and temperatures centered around the critical point. 

The ionic concentrations on the other hand were sampled uniformly while explicitly considering pure water, single-salt, and mixed-salts conditions. The synthetic dataset contains 150,000 samples for pure water, 10,000 samples for single salt solutions, and 10,000 samples for mixed salt solutions. Concentration limits were imposed on both the sigle- and multi-salt solutions.

The training process involves two steps: 1) training the model to explicitly predict the intermediary variables \citet{Duan2006} model, and 2) fine-tuning the model on experimental data for \COtwo solubility. During this second step, the different intermediary outputs are combined following \citet{Duan2003} model.

During the first training stage, the model was trained for 300 epochs with an early stopping criterion of 20 epochs without improvement in the validation loss. The model was trained to predict the intermediary variables only. During the second training stage, the model was fine-tuned on a combination of synthatic and experimental data. Experimental data points were assigned higher weights compared to synthetic data, to emphasize their importance. The synthetic data generated from \citet{Duan2006} exhibit a relatively higher error at high salinities, so we put a higher weight on experimental data points corresponding to higher salt concentrations. Adam optimizer was used for the training.

\section{Results}
The performance of the physics-informed multitask learning model was evaluated in two stages, mirroring the training process. The initial stage focused on the model's ability to replicate Duan \& Sun model, while the final stage assessed the fine-tuned model's predictive accuracy against experimental test data.

The first training stage aimed to embed the physical relationships from the \citet{Duan2006} model into our neural network architecture. The model was trained on the large synthetic dataset to predict the intermediary thermodynamic variables and the final \COtwo solubility. As detailed in Table~\ref{table:first_training_stage_results}, the model achieved a high degree of fidelity in this task. The R\textsuperscript{2} values for all outputs were greater than 0.998, and the AAD for each intermediary variable was exceptionally low. This result confirms that the multitask architecture successfully learned the functions governing \COtwo dissolution as described by the Duan and Sun model, resulting in a physically-consistent base prior to fine-tuning on experimental data.

\begin{table}[H]
	\centering
	\caption{Performance metrics for the first training stage, where the model was trained to replicate the \citet{Duan2006} model on synthetic data. AAD denotes Average Absolute Deviation.}
	\label{table:first_training_stage_results}
	\begin{tabular}{l|cc}
		\hline
		Output Variable        & AAD    & R\textsuperscript{2} \\
		\hline
		$\log(\phi)$           & 0.0046 & 0.9988 \\
		$\mu_{\textrm{\COtwo}}^{l(0)}/RT$ & 0.0059 & 0.9999 \\
		$\log(\gamma)$           & 0.0078 & 0.9995 \\
		\hline
		Dissolved CO\textsubscript{2} (molality)      & 0.0203 & 0.9982 \\
		\hline
	\end{tabular}
\end{table}

Following the pre-training, the model was fine-tuned using a combination of synthetic and experimental data. Its final performance was evaluated on a held-out test set and compared against three benchmarks: PHREEQC, the baseline blackbox model, and Duan \& Sun model. The results of this comparison are summarized in Table~\ref{table:models_metrics}.

\begin{figure}[H]
	\centering
	\includegraphics[width=0.5\textwidth]{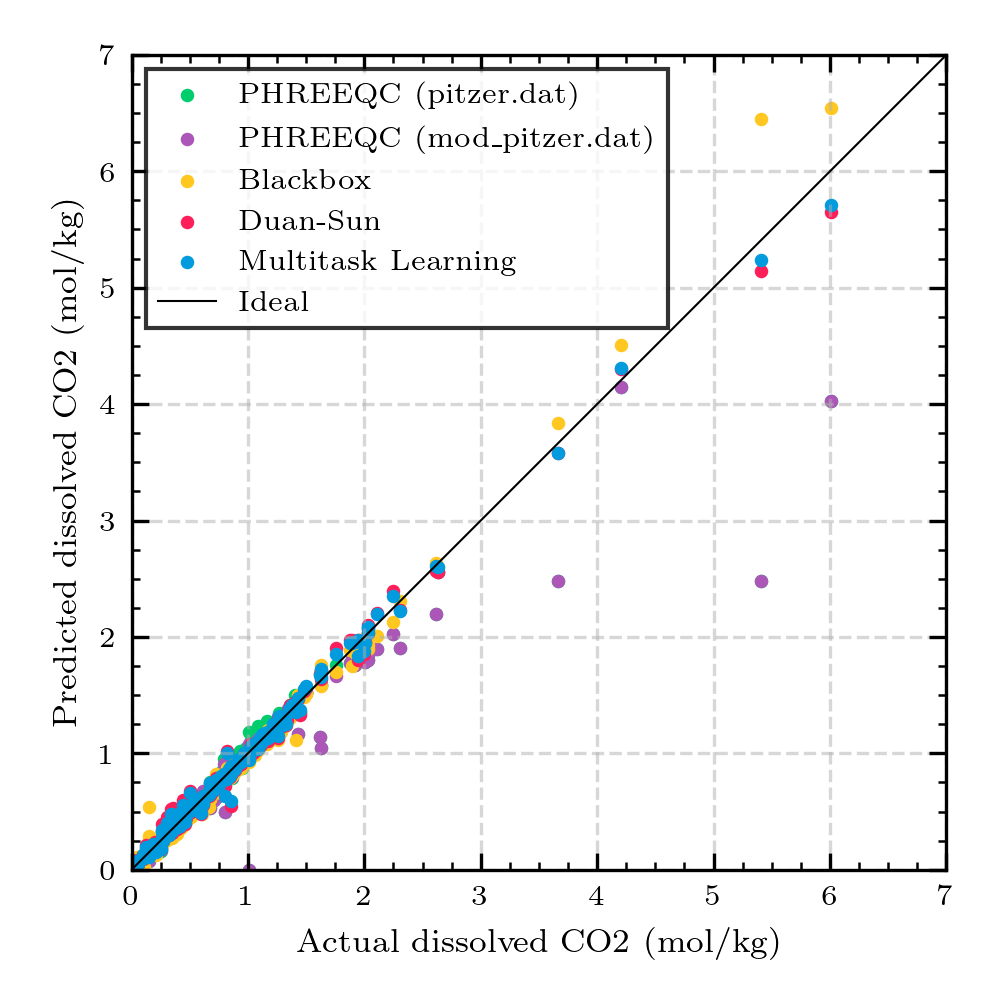}
	\caption{Predicted versus experimental \COtwo molality for the multitask learning model on the test set. The dashed line represents the line of parity (y=x).}
	\label{fig:predictions-vs-actual}
\end{figure}

Our proposed multitask learning model demonstrates superior performance over all benchmarks, achieving the lowest AAD of 0.02877 and the highest R\textsuperscript{2} value of 0.9952. Compared to the PHREEQC baseline, our model shows a remarkable 72.87\% improvement in AAD. Critically, the multitask model also outperforms the blackbox model, indicating that the physics-informed architecture and the pre-training step provide a distinct advantage over a simple, data-driven approach, while using the same experimental training data. Figure~\ref{fig:predictions-vs-actual} illustrates the parity plot for predictions from different models. The parity plot also shows that all models perform reasonably well, with PHREEQC error being aggravated mainly due to a few outliers corresponding mostly to higher pressure data points.

\begin{figure}[H]
	\centering
	\includegraphics[width=0.5\textwidth]{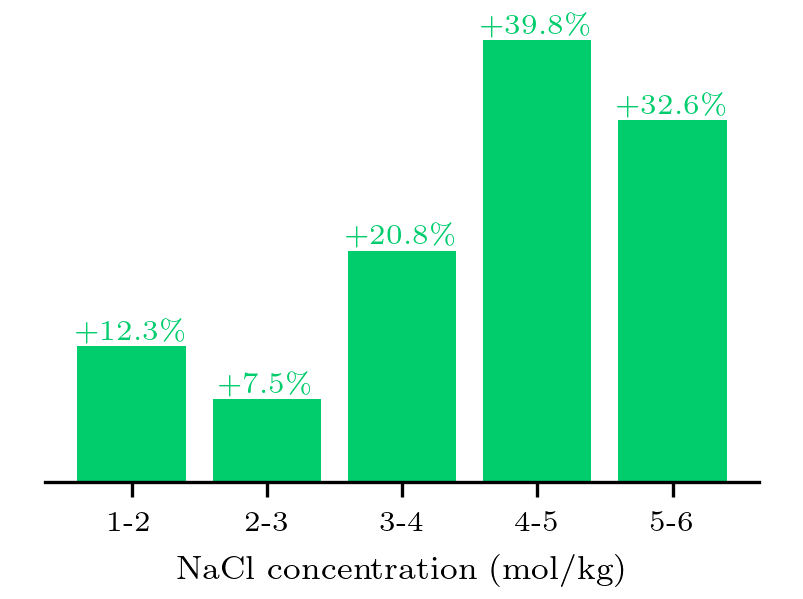}
	\caption{Percent improvement in AAD of the multitask learning model relative to the \citet{Duan2006} model as a function of NaCl molality. The improvement is most pronounced at higher salinities.}
	\label{fig:error-variance-with-salinity}
\end{figure}

The fine-tuned model exhibits a 14.4\% improvement in AAD over the Duan \& Sun model it was pre-trained to replicate. This highlights the success of the fine-tuning stage in correcting for the inherent errors of the source model by incorporating real-world experimental data. This is a good but not a significant improvement on a first look. However, when we break down the AAD by salinity, our proposed model achieves significantly lower AAD at higher salinities compared to \citet{Duan2006}; a 20\%--40\% lower AAD in NaCl brines for example at higher salinities, emphasizing (Figure~\ref{fig:error-variance-with-salinity}).

\begin{table}[H]
	\centering
	\caption{Evaluation of different models on the experimental test set. The percentage improvement in AAD is calculated relative to the PHREEQC baseline.}
	\label{table:models_metrics}
	\begin{tabular}{l|ccc}
		\hline
		Model               & AAD    & R\textsuperscript{2}  & \% Imp. \\
		\hline
		PH (\texttt{pitzer})             & 0.0709  & 0.8790 & Baseline\\
		PH (\texttt{mod\_pitzer})             & 0.0564  & 0.8817 & 25.74\%\\
		Blackbox            & 0.04641  & 0.9827 & 34.54\%\\
		\citet{Duan2006}    & 0.03361  & 0.9931 & 53.60\%\\
		Multitask Learning  & 0.02877 & 0.9952 & 59.41\%\\
		\hline
	\end{tabular}
\end{table}

The three models, namely PHREEQC, Duan \& Sun, and our proposed physics-informed neural network, present a classic trade-off between specialized accuracy and general versatility. PHREEQ is a comprehensive geochemical simulator that offers the greatest versatility for modeling complex water-rock-gas interactions but is the least accurate for this specific solubility task. In contrast, the Duan--Sun model is a simple and highly accurate physical equation but is rigid and limited exclusively to \COtwo--brine systems. While its activity model is very accurate, it starts to deviate from experimental measurements at higher salinities. The Multitask Learning model represents the best of both worlds, achieving the highest accuracy by fine-tuning physical principles with real-world data, resulting in a fast, superiorly predictive, and extensible model ideal for dynamic simulations and accommodating future data or gas impurities.
\section{Conclusion}
This work provided a comprehensive review of \COtwo solubility in brines, covering experimental measurement techniques, physical models, and numerical simulation approaches. We also examined the key factors influencing \COtwo solubility, including pressure, temperature, water chemistry, and the presence of impurities.

A primary contribution of this research is the development of a new, temperature-dependent set of Pitzer interaction parameters for PHREEQC's \texttt{pitzer.dat} thermodynamic database. These new parameters demonstrated significant accuracy improvements over default settings when benchmarked against experimental data. Importantly, while fitted using single-salt solubility data, they also exhibited strong generalization to multi-salt systems, achieving lower errors on data not used for fitting.

Furthermore, we introduced a novel physics-informed machine learning model for predicting \COtwo solubility in brines. This model leverages fundamental physical knowledge and thermodynamic principles to create a custom architecture. Trained on both synthetic and experimental data, it achieved superior accuracy compared to existing state-of-the-art models.

\bibliography{misc/main} 

\end{multicols}

\end{document}